


\font\twelverm=cmr12
\font\twelvei=cmmi12
\font\twelvesy=cmsy10 scaled 1200
\font\twelvebf=cmbx12
\font\twelvett=cmtt12
\font\twelveit=cmti12
\font\twelvesl=cmsl12
\font\twelvesc=cmcsc10 scaled 1200
\font\twelvess=cmss12
\font\twelvessi=cmssi12
\font\twelvebmit=cmmib10 scaled 1200
\font\twelvebsy=cmbsy10 scaled 1200

\font\elevenrm=cmr10 scaled 1095
\font\eleveni=cmmi10 scaled 1095
\font\elevensy=cmsy10 scaled 1095
\font\elevenbf=cmbx10 scaled 1095
\font\eleventt=cmtt10 scaled 1095
\font\elevenit=cmti10 scaled 1095
\font\elevensl=cmsl10 scaled 1095
\font\elevensc=cmcsc10 scaled 1095
\font\elevenss=cmss10 scaled 1095
\font\elevenssi=cmssi10 scaled 1095
\font\elevenbmit=cmmib10 scaled 1095
\font\elevenbsy=cmbsy10 scaled 1095

\font\tensc=cmcsc10
\font\tenss=cmss10
\font\tenssi=cmssi10
\font\tenbmit=cmmib10
\font\tenbsy=cmbsy10

\font\ninerm=cmr9    \font\eightrm=cmr8    \font\sixrm=cmr6
\font\ninei=cmmi9    \font\eighti=cmmi8    \font\sixi=cmmi6
\font\ninesy=cmsy9   \font\eightsy=cmsy8   \font\sixsy=cmsy6
\font\ninebf=cmbx9   \font\eightbf=cmbx8   \font\sixbf=cmbx6
\font\ninett=cmtt9   \font\eighttt=cmtt8
\font\nineit=cmti9   \font\eightit=cmti8   
\font\ninesl=cmsl9   \font\eightsl=cmsl8
\font\niness=cmss9   \font\eightss=cmss8
\font\ninessi=cmssi9 \font\eightssi=cmssi8


\skewchar\twelvei='177    \skewchar\eleveni='177     \skewchar\ninei='177
\skewchar\eighti='177     \skewchar\sixi='177
\skewchar\twelvesy='60    \skewchar\elevensy='60     \skewchar\ninesy='60
\skewchar\eightsy='60     \skewchar\sixi='60


\catcode`@=11
\newskip\ttglue
\parindent=3em

\def\twelvept{\def\rm{\fam0\twelverm}
   \textfont0=\twelverm \scriptfont0=\ninerm \scriptscriptfont0=\sevenrm%
   \textfont1=\twelvei  \scriptfont1=\ninei  \scriptscriptfont1=\seveni%
   \textfont2=\twelvesy \scriptfont2=\ninesy \scriptscriptfont2=\sevensy%
   \textfont3=\tenex    \scriptfont3=\tenex  \scriptscriptfont3=\tenex%
   \textfont\itfam=\twelveit   \def\it{\fam\itfam\twelveit}%
   \textfont\slfam=\twelvesl   \def\sl{\fam\slfam\twelvesl}%
   \textfont\ttfam=\twelvett   \def\tt{\fam\ttfam\twelvett}%
   \textfont\bffam=\twelvebf   \scriptfont\bffam=\ninebf%
      \scriptscriptfont\bffam=\sevenbf   \def\bf{\fam\bffam\twelvebf}%
   \def\oldstyle{\fam1 \twelvei}%
   \tt \ttglue=.5em plus.25em minus.15em%
   \normalbaselineskip=13pt plus.5pt minus.5pt%
   \def\doublespacing{\baselineskip=26pt plus.5pt minus 1pt}%
   \def\spaceandahalf{\baselineskip=19.5pt plus.5pt minus .5pt}%
   \setbox\strutbox=\hbox{\vrule height9pt depth4pt width0pt}%
   \let\sc=\twelvesc   \let\big=\tenbig%
   \let\ss=\twelvess   \let\ssi=\twelvessi%
   \let\bmit=\twelvebmit\let\bsy=\twelvebsy%
   \normalbaselines\rm}

\def\elevenpt{\def\rm{\fam0\elevenrm}
   \textfont0=\elevenrm \scriptfont0=\eightrm \scriptscriptfont0=\sixrm%
   \textfont1=\eleveni  \scriptfont1=\eighti  \scriptscriptfont1=\sixi%
   \textfont2=\elevensy \scriptfont2=\eightsy \scriptscriptfont2=\sixsy%
   \textfont3=\tenex    \scriptfont3=\tenex  \scriptscriptfont3=\tenex%
   \textfont\itfam=\elevenit   \def\it{\fam\itfam\elevenit}%
   \textfont\slfam=\elevensl   \def\sl{\fam\slfam\elevensl}%
   \textfont\ttfam=\eleventt   \def\tt{\fam\ttfam\eleventt}%
   \textfont\bffam=\elevenbf   \scriptfont\bffam=\eightbf%
      \scriptscriptfont\bffam=\sixbf   \def\bf{\fam\bffam\elevenbf}%
   \def\oldstyle{\fam1 \eleveni}%
   \tt \ttglue=.5em plus.25em minus.15em%
   \normalbaselineskip=12pt plus.5pt minus.5pt%
   \def\doublespacing{\baselineskip=24pt plus.5pt minus1pt}%
   \def\spaceandahalf{\baselineskip=18pt plus.5pt minus .5pt}%
   \setbox\strutbox=\hbox{\vrule height8.5pt depth3.5pt width0pt}%
   \let\sc=\elevensc   \let\big=\tenbig
   \let\ss=\elevenss   \let\ssi=\elevenssi%
   \let\bmit=\elevenbmit\let\bsy=\elevenbsy%
   \normalbaselines\rm}

\def\tenpt{\def\rm{\fam0\tenrm}
   \textfont0=\tenrm \scriptfont0=\sevenrm \scriptscriptfont0=\fiverm%
   \textfont1=\teni  \scriptfont1=\seveni  \scriptscriptfont1=\fivei%
   \textfont2=\tensy \scriptfont2=\sevensy \scriptscriptfont2=\fivesy%
   \textfont3=\tenex \scriptfont3=\tenex   \scriptscriptfont3=\tenex%
   \textfont\itfam=\tenit   \def\it{\fam\itfam\tenit}%
   \textfont\slfam=\tensl   \def\sl{\fam\slfam\tensl}%
   \textfont\ttfam=\tentt   \def\tt{\fam\ttfam\tentt}%
   \textfont\bffam=\tenbf   \scriptfont\bffam=\sevenbf%
      \scriptscriptfont\bffam=\fivebf   \def\bf{\fam\bffam\tenbf}%
   \def\oldstyle{\fam1 \teni}%
   \tt \ttglue=.5em plus.25em minus.15em%
   \normalbaselineskip=11pt plus.5pt minus.5pt%
   \def\doublespacing{\baselineskip=22pt plus.5pt minus 1pt}%
   \def\spaceandahalf{\baselineskip=16.5pt plus.5pt minus .5pt}%
   \setbox\strutbox=\hbox{\vrule height8.5pt depth3.5pt width0pt}%
   \let\sc=\tensc   \let\big=\tenbig%
   \let\ss=\tenss   \let\ssi=\tenssi%
   \let\bmit=\tenbmit\let\bsy=\tenbsy%
   \normalbaselines\rm}

\def\ninept{\def\rm{\fam0\ninerm}
   \textfont0=\ninerm \scriptfont0=\sixrm \scriptscriptfont0=\fiverm%
   \textfont1=\ninei  \scriptfont1=\sixi  \scriptscriptfont1=\fivei%
   \textfont2=\ninesy \scriptfont2=\sixsy \scriptscriptfont2=\fivesy%
   \textfont3=\tenex  \scriptfont3=\tenex \scriptscriptfont3=\tenex%
   \textfont\itfam=\nineit   \def\it{\fam\itfam\nineit}%
   \textfont\slfam=\ninesl   \def\sl{\fam\slfam\ninesl}%
   \textfont\ttfam=\ninett   \def\tt{\fam\ttfam\ninett}%
   \textfont\bffam=\ninebf   \scriptfont\bffam=\sixbf%
      \scriptscriptfont\bffam=\fivebf   \def\bf{\fam\bffam\ninebf}%
   \def\oldstyle{\fam1 \ninei}%
   \tt \ttglue=.5em plus.25em minus.15em%
   \normalbaselineskip=10pt plus.5pt minus.5pt%
   \def\doublespacing{\baselineskip=20pt plus.5pt minus1pt}%
   \def\spaceandahalf{\baselineskip=15pt plus.5pt minus .5pt}%
   \setbox\strutbox=\hbox{\vrule height8pt depth3pt width0pt}%
   \let\sc=\sevenrm   \let\big=\ninebig%
   \let\ss=\niness    \let\ssi=\ninessi%
   \normalbaselines\rm}

\def\eightpt{\def\rm{\fam0\eightrm}
   \textfont0=\eightrm \scriptfont0=\sixrm \scriptscriptfont0=\fiverm%
   \textfont1=\eighti  \scriptfont1=\sixi  \scriptscriptfont1=\fivei%
   \textfont2=\eightsy \scriptfont2=\sixsy \scriptscriptfont2=\fivesy%
   \textfont3=\tenex   \scriptfont3=\tenex \scriptscriptfont3=\tenex%
   \textfont\itfam=\eightit   \def\it{\fam\itfam\eightit}%
   \textfont\slfam=\eightsl   \def\sl{\fam\slfam\eightsl}%
   \textfont\ttfam=\eighttt   \def\tt{\fam\ttfam\eighttt}%
   \textfont\bffam=\eightbf   \scriptfont\bffam=\sixbf%
      \scriptscriptfont\bffam=\fivebf   \def\bf{\fam\bffam\eightbf}%
   \def\oldstyle{\fam1 \eighti}%
   \tt \ttglue=.5em plus.25em minus.15em%
   \normalbaselineskip=9pt plus.5pt minus.5pt%
   \def\doublespacing{\baselineskip=18pt plus.5pt minus1pt}%
   \def\spaceandahalf{\baselineskip=13.5pt plus.5pt minus .5pt}%
   \setbox\strutbox=\hbox{\vrule height7pt depth2pt width0pt}%
   \let\sc=\sixrm   \let\big=\eightbig%
   \let\ss=\eightss \let\ssi=\eightssi%
   \normalbaselines\rm}

\def\tenbig#1{{\hbox{$\left#1\vbox to8.5pt{}\right.\n@space$}}}
\def\ninebig#1{{\hbox{$\textfont0=\tenrm\textfont2=tensy
   \left#1\vbox to7.25pt{}\right.\n@space$}}}
\def\eightbig#1{{\hbox{$\textfont0=\ninerm\textfont2=ninesy
   \left#1\vbox to6.5pt{}\right.\n@space$}}}

\let\singlespacing=\normalbaselines


\def\lft#1{{#1}\hfill}
\def\ctr#1{\hfill{#1}\hfill}


\def\today{\ifcase\month\or
   January\or February\or March\or April\or May\or June\or
   July\or August\or September\or October\or November\or December\fi
   \space\number\day, \number\year}

\def\sciday{\number\day
   \space\ifcase\month\or
   January\or February\or March\or April\or May\or June\or
   July\or August\or September\or October\or November\or December\fi
   \space\number\year}

\newcount\tyme
\newcount\hour
\newcount\minute

\def\amorpm{a.m.}
\def\tod{\gettime\number\hour:\ifnum\minute<10{}0\fi\number\minute\space\amorpm}

\def\gettime{\tyme=\time
   \divide \tyme by 60
   \hour=\tyme
   \ifnum\hour=12\gdef\amorpm{p.m.}\fi
   \ifnum\hour=0 \advance \hour by  12\fi
   \ifnum\hour>12\advance \hour by -12\gdef\amorpm{p.m.}\fi
   \multiply \tyme by 60
   \advance \time by -\tyme
   \minute=\time
   \advance \time by \tyme}



\def\underule#1{$\setbox0=\hbox{#1} \dp0=\dp\strutbox
    \m@th \underline{\box0}$}


\def\narrow{\advance\leftskip by3em \advance\rightskip by3em}
\def\wide{\advance\leftskip by-3em \advance\rightskip by-3em}


\def\alph#1{\ifcase#1\or a\or b\or c\or d\or e\or f\or g\or h\or i\or j\or
   k\or l\or m\or n\or o\or p\or q\or r\or s\or t\or u\or v\or w\or x\or
   y\or z\fi}

\def\deg{\ifmmode^\circ\else$^\circ$\fi}


\def\applt{\mathrel{\mathpalette\@versim<}}
\def\appgt{\mathrel{\mathpalette\@versim>}}
\def\@versim#1#2{\lower2pt\vbox{\baselineskip0pt \lineskip-.5pt
   \ialign{$\m@th#1\hfil##\hfil$\crcr#2\crcr\sim\crcr}}}


{\catcode`p=12 \catcode`t=12 \gdef\\#1pt{#1}}
\let\getfactor=\\
\def\kslant#1{\kern\expandafter\getfactor\the\fontdimen1#1\ht0}
\def\vector#1{\ifmmode\setbox0=\hbox{$#1$}%
    \setbox1=\hbox{\the\scriptscriptfont1\char'52}%
	\dimen@=-\wd1\advance\dimen@ by\wd0\divide\dimen@ by2%
    \rlap{\kslant{\the\textfont1}\kern\dimen@\raise\ht0\box1}#1\fi}

\tenpt\singlespacing

\vsize=9truein
\hsize=6.5truein
\hoffset=1truein
\voffset=1truein
\twelvept
\def\slash#1{\setbox0=\hbox{$#1$}#1\hskip-\wd0\hbox to\wd0{\hss\sl/\/\hss}}
\overfullrule=0pt
\nopagenumbers
\rightline {ISU-NP-93-02, KU-HEP-93-29}
\bigskip

\centerline{\bf Isospin Multiplet Structure in Ultra--Heavy Fermion Bound
States}
\vskip .5in
\centerline{{\bf Pankaj Jain},$^a$ {\bf Alan J. Sommerer},$^b$ {\bf Douglas
W. McKay},$^a$}
\centerline{{\bf J.R. Spence},$^b$ {\bf J.P. Vary},$^b$ and {\bf Bing--Lin
Young},$^b$ }
\bigskip
\item{$^a$} Department of Physics and Astronomy, University of Kansas,
Lawrence, Kansas 66045--2151
\smallskip
\item{$^b$} Department of Physics and Astronomy and Ames Laboratory, Iowa State
University, Ames, Iowa 50011.

\vskip .5in
\centerline{\bf Abstract}
\medskip
\spaceandahalf
The coupled Bethe--Salpeter bound state equations for a $Q\bar Q$ system, where
$Q=(U,D)$ is a degenerate, fourth generation, super--heavy quark doublet, are
solved in several ladder approximation models. The exchanges of gluon, Higgs
and Goldstone modes in the standard model are calculated in the ultra--heavy
quark limit where weak $\gamma, W^\pm$ and $Z^0$ contributions are negligible.
A natural $I=0$ and $I=1$ multiplet pattern is found, with large splittings
occuring between the different weak iso--spin states when $M_Q$, the quark
masses, are larger than values in the range $0.4 TeV<M_Q<0.8 TeV$, depending on
which model is used.
\medskip
Consideration of ultra--heavy quark lifetime constraints and $U-D$ mass
splitting constraints are reviewed to establish the plausibility of lifetime
and mass degeneracy requirements assumed for this paper.
\vfill
\centerline {Typeset in \TeX\ by Reta Solwa}
\vfill\eject

\footline={\hss\elevenrm\folio\hss}
\pageno=2

\noindent {\bf 1. Introduction}
\medskip
The prospect of SSC and LHC experiments to probe TeV phenomena has stimulated a
wide variety of ideas about the physics at these energies. The standard model
itself, expanded to a super--heavy fourth
generation, could display some striking new effects.
There is a variety of ``heavy--neutrino" fourth generation models in the
literature,$^1$ and we adopt the attitude that some model of this type with the
Higgs
mechanism providing the fermion and gauge boson mass generation represents
nature at 0.5-1.0 TeV.
The effect that we
explore in this
paper is the strong Yukawa coupling of the Higgs sector to the fourth
generation and the bound state spectrum in the heavy quark--anti--quark system
($Q\bar Q$) that is generated by this strong coupling.$^{2,3}$ There have been
a number
of heavy fermion physics studies since the Higgs mechanism and the standard
model were proposed.$^{4,5}$ Like the heavy Higgs limit studies, the
indications are
that perturbative calculations break down and some new, strong coupling
features of the standard model and/or totally new physics should emerge at the
0.5--1.0 TeV energy ranges.
\medskip
In a previous work,$^3$ we described some new, deep--binding features of the
$0^-$ ground state of a single, super--heavy $Q\bar Q$ system that were driven
by the dominant, attractive Higgs scalar exchange plus weaker, but in some
cases still significant, gluon exchange. The general feature was that a very
strong binding compared to QCD alone was generally exhibited for $M_Q>0.4-0.5$
TeV in all of the model calculations that we tested.
\medskip
In this paper, we expand significantly on this theme by doing a coupled channel
analysis of a heavy quark doublet, ($U,D$), including the Goldstone boson
exchanges as well as the Higgs exchanges, and we discuss the $s$--channel
exchange as well as the $t$--channel exchange contributions to the
Bethe--Salpeter equation kernel. Here we include the results of calculations of
the $0^+$ and $1^-$ masses in addition to the $0^-$ ground state case analyzed
previously.
\medskip
Because of the weak iso--doublet structure of the super--heavy ($U,D$) system
the bound states decomposes naturally into weak iso-triplet and weak
iso--singlet states when $M_U\simeq M_D$ and $M_Z\simeq M_W$. We find that the
interplay between the scalar Higgs exchange and the (opposite sign)
pseudoscalar Goldstone boson (longitudinal $W$ and $Z$) exchange results in a
clear iso--triplet/iso--singlet mass splitting in all of the channels in all
model calculations we used. This signature is quite distinct for the fourth
generation fermion doublet, since it must be approximately degenerate according
to
present constraints from $W/Z$ physics.
\medskip
In the following section, we outline the coupled--channel Bethe--Salpeter
formalism
used to calculate bound state masses. In Sec. 3 we present results of the
calculations in the different approximation schemes used and comment on the
main features. In Sec. 4 we show that requiring meta--stability of fourth
generation quarks puts only weak constraints on mixing with a
lighter generation.
We also review the evidence that a fourth generation of
quarks, if they exist, would have to be approximately degenerate in mass. In
Sec. 5 we summarize and conclude.  In Appendix A, we describe the subtraction
used to regulate the Bethe--Saleter equations.
In
Appendix B, we give a short analysis of the running of the Yukawa coupling and
its
relevance to the calculations presented.
\bigskip
\noindent {\bf 2. Strong Yukawa Binding of Ultra--Heavy Fermion Weak
Doublets: Bethe--Salpeter Equation Formalism Including Goldstone Boson
Exchange}
\medskip
If there are fourth and higher generations of fermion weak isospin doublets,
the fractional mass--splitting within these ultra--heavy
doublets$^6$ is constrained by the
standard, three
generation model's remarkable survival after two decades of stringent
experimental tests.$^7$ A small mass--splitting, or
near
degeneracy, could lead to a classic isospin--multiplet pattern for the bound
state spectrum, and we show in this section that this is indeed the case when
the Yukawa couplings of the Higgs and
Goldstone--bosons$^8$ are incorporated into the bound state equations -- ladder
approximation,
Bethe--Salpeter equations in our study. We assume that the usual mass
generation
mechanism operates in the fourth generation sector as well, so the ultra heavy
fermion doublets will have strong Yukawa couplings to the standard model Higgs
and Goldstone modes. These Yukawa couplings, scalar for the Higgs boson and
pseudoscalar for Goldstone bosons, become the dominant ones when the quark
masses reach 400--500 GeV (though the pseudoscalar exchanges actually become
competitive only in the tight binding, highly relativistic regime. They
decouple in the weak--binding limit).
\medskip
We develop below a set--up for the two, coupled--channel Bethe-Salpeter
equations of a quark doublet--antidoublet that interact through gluon and
weak--sector Higgs and Goldstone boson exchange.
\medskip
Let us consider a doublet of heavy quarks, ($U,D$), interacting in the Feynman
gauge by exchange of
gluons, $g,$ a Higgs boson, $H$, and Goldstone bosons $\chi^\pm,\chi^0$. The
photon and $W^\pm$ and $Z$ exchanges are always weak in the gauges chosen, have
been checked and
found to
give insignificant contributions to the binding, and are not included in our
development. It is convenient to adopt Feynman gauge in the ladder
approximation
(perturbative vertices, no crossed graphs), which we will use throughout this
work. There are four quark--antiquark channels: $\bar UU,\; \bar UD,\; \bar DU$
and $\bar DD$. These channels become coupled when the charged Goldstone boson
exchanges are included, as illustrated in Fig. 1, where
the  Bethe--Salpeter equations are shown graphically. As discussed below,
these equations admit a subtraction at fixed $q$ that eliminates the
divergent, $q$--independent, annihilation graphs which enter the scalar and
pseudoscalar bound state channels.$^9$ The corresponding gluon annihilation
graph
does not require consideration because we study only color singlet bound
states.
\medskip
The coupled--channel, doublet quark--antiquark bound state equations are shown
pictorially in Fig. 1 and displayed in Eqs. (2.1) and (2.2) below.
The subtracted, decoupled versions of these equations,
are presented later in this section, Eqs. (2.3a and 2.3b). Let us display, in
shorthand
notation, the structure of the ladder approximation, momentum -- space
Bethe--Salpeter equations for the bound state amplitudes $\chi(q,P)$ in Feynman
gauge.
\medskip
$\bar U-U$ Channel:
$$\eqalign{S^{-1}(q_+)&\chi_{_{UU}}(q,P)S^{-1}(q_-)=\int {d^4k\over
(2\pi)^4}\bigg\lbrack -\gamma_\mu\chi_{_{UU}}(k,P)\gamma_\nu G^{\mu\nu}(k-q)\cr
&-{g_2^2\over 4}{m^2\over M^2_W}\chi_{_{UU}}(k,P){i\over
(k-q)^2-M_H^2}+{g_2^2\over
4}{m^2\over M_W^2}\gamma_5\chi_{_{UU}}(k,P)\gamma_5 {i\over (k-q)^2-M_Z^2}\cr
&+{g_2^2\over 2}{m^2\over M_W^2}\gamma_5\chi_{_{DD}}(k,P)\gamma_5{i\over
(k-q)^2-M_W^2}\cr
&+{g_2^2\over 4}{m^2\over M_W^2} {i\over P^2-M_Z^2}\gamma_5
Tr\Biggl(\biggl(\chi_{_{UU}}- \chi_{_{DD}}\biggr)\gamma_5\Biggr)+{g^2_2\over
4}{m^2\over
M_W^2}{i\over
P^2-M_H^2}Tr\biggl(\chi_{_{UU}}+\chi_{_{DD}} \biggr)
\bigg\rbrack\; \; .\cr}\eqno (2.1)$$
$\bar D-D$ Channel: interchange $U\leftrightarrow D$ in (2.1).
\smallskip
\noindent $\bar U-D$ Channel:
$$\eqalign{S^{-1}(q_+)\chi_{_{UD}}(q,P)&S^{-1}(q_-)=\int {d^4k\over
(2\pi)^4}\bigg\lbrack -\gamma_\mu\chi_{_{UD}}(k,P)\gamma_\nu G^{\mu\nu}(k-q)\cr
&-{g_2^2\over 4}{m^2\over
M^2_W}\chi_{_{UD}}(k,P){i\over(k-q)^2-M_H^2}-{g_2^2\over
4}{m^2\over M_W^2}\gamma_5\chi_{_{UD}}(k,P)\gamma_5{i\over (k-q)^2-M_Z^2}\cr
&+{g_2^2\over 2}{m^2\over M_W^2} {i\over P^2-M_W^2}\gamma_5
Tr\biggl(\chi_{_{UD}}(k,P)\gamma_5\biggr)\bigg\rbrack\; \; .\cr}\eqno (2.2)$$
$\bar D-U$ Channel: interchange $U\leftrightarrow D$ in Eq. (2.2).
\medskip
In Eqs. (2.1) and (2.2), $\chi_{_{UU}}(q,P)$ etc., represent the
Bethe--Salpeter
amplitudes for the $\bar UU$ etc., systems with relative 4--momentum $q$ and
total 4--momentum $P$. The quark masses, $m$, are assumed to be degenerate,
$q_\pm\equiv q\pm {P\over 2}$ and $S^{-1}(q_\pm)$ are the inverse quark
propagators with momenta $q_\pm$. Here $G^{\mu\nu}(k-q)$ represents the gluon
propagator, assumed to behave like $(k-q)^{-2}$ as $\mid k-q\mid\to \infty$.
The SU(2) coupling constant is denoted by $g_2$.

\medskip
One may decouple the $\bar UU$ and $\bar DD$ equations by taking the
combinations $(\bar UU\pm \bar DD)/\sqrt{2}$, and the resulting Bethe--Salpeter
equations, subtracted as described in Appendix A,
for the bound state wave--functions $\chi_{_{UD}}(q,P),\; \chi_-(q,P)$
and $\chi_+(q,P)$ read as follows $(\chi_{_{DU}}(q,P)$ obeys the same equations
as
$\chi_{_{UD}})$:
$$\eqalign{S^{-1}(q_+)\chi_\pm(q,P)S^{-1}(q_-)&=\int{d^4k\over
(2\pi)^4}\bigg\lbrace -\gamma_\mu \chi_\pm (k,P)\gamma_\nu G^{\mu\nu}(k-q)\cr
-i{m^2g^2_2\over 4M^2_W} {\chi_\pm(k,P)\over(k-q)^2-M^2_H}&+{i\over
4}{m^2g^2_2\over M_{W}^2} {\gamma_5\chi_\pm(k,P)\gamma_5\over (k-q)^2-M_Z^2}
\mp {i\over 2} {m^2g_2^2\over M_{W}^2} {\gamma_5\chi_\pm(k,P)\gamma_5\over
(k-q)^2-M_{W}^2}\bigg\rbrace\cr}\eqno (2.3a)$$ and
$$\eqalign{S^{-1}(q_+)\chi_{_{UD}}(q,P)S^{-1}(q_-)&=\int{d^4k\over
(2\pi)^4}\bigg\lbrace -\gamma_\mu \chi_{_{UD}} (k,P)\gamma_\nu
G^{\mu\nu}(k-q)\cr
-i{m^2g^2_2\over 4M_{W}^2} {\chi_{_{UD}}(k,P)\over(k-q)^2-M_{H}^2}&-
i{m^2g^2_2\over 4M_{W}^2} {\gamma_5\chi_{_{UD}}\gamma_5\over (k-q)^2-M_Z^2}
\bigg\rbrace\cr}\eqno (2.3b)$$ The quark
mass $m=m_U=m_D$ is taken to be the same for $U$ and $D$. Experimental support
for this assumption is reviewed in Sec. 4.
\medskip
The important features of Eq. (2.3) are
\smallskip
\item{(i)} $\chi_-,\; \chi_{_{UD}}$ and $\chi_{_{DU}}$ obey the {\it same
equation}
when $M_Z=M_W$, and they form an isospin triplet of states.
\smallskip
\item{(ii)} The Goldstone boson contributions to this
$\chi_-,\chi_{_{UD}},
\chi_{_{DU}}$ triplet are, like the gluon and Higgs boson
contributions {\it all attractive} in the {\it ground state}, $0^-$ system,
(see Sec. 3 below)
\smallskip
\item{(iii)} the $\chi_+$ state obeys a different equation whose
Goldstone--boson contributions are all {\it repulsive} in the {\it ground
state}, $0^-$ system.
\smallskip
\noindent Eq. (2.3) and the fact that pseudoscalar exchange becomes
comparable to
scalar for relativistic systems combine to produce a clear splitting between
the singlet and triplet of ground state bosons in the strong binding limit. The
triplet is below the singlet in mass for the $0^-$ system. This is very
reminiscent of a
``$\pi-\eta$" situation, except that this appears to be purely dynamical with
no spontaneous symmetry breaking, bound state Goldstone phenomenon at work.
\medskip
The  iso--singlet, iso--triplet situation is reversed in the $0^+$ and $1^-$
spin--parity channels, as we describe in the next section.
\bigskip
\noindent {\bf 3. Results of Relativistic Bound State Calculations of
Iso--Singlet and Iso--Triplet Masses}
\medskip
We employ two different approaches to solving the ladder approximation,
Bethe--Salpeter equations for the $J^P=0^-$ ground states of the weak iso--spin
singlet and weak iso--spin triplet $Q\bar Q$ system. We also report results in
the same approximations for the $J^P=1^-$ and $0^+$ states. The principle
feature that
emerges, as anticipated in the preceding section, is the splitting between the
iso--singlet and iso--triplet states due to the pseudoscalar exchange (in
Feynman gauge): repulsion in the iso--singlet case and attraction in the
iso--triplet
case in the ground state, $0^-$, channel. The iso--triplet pseudoscalar, ground
state mass solutions are found to fall to zero for high enough
quark masses in all of the calculations. The reverse is true in the $0^+$ and
$1^-$ channels, and the isoscalar masses fall to zero at high enough quark
masses in these states.$^{10}$
\medskip
We evaluate the bound state energies (masses) using both the covariant kernel,
$$K(q)=g^2{1\over q^2-M^2}\; \; ,\eqno (3.1a)$$ which we refer to as the
covariant gauge ladder approximation, and the instantaneous approximation to
this kernel
$$K(\mid\vec q\mid^2)=-g^2{1\over \mid\vec q\mid^2+M^2}\; \;.\eqno (3.1b)$$ We
have suppressed any reference to gauge--dependent spin structure in (3.1), and
designate generic coupling constants and masses as $g$ and $M$, respectively.
The instantaneous approximation, so called because the coordinate space
potentials are instantaneous, reduces the problem to three dimensions, yielding
what are referred to as Salpeter's equations. These are coupled equations for
positive and negative frequency amplitudes. One can solve the coupled system or
set the negative frequency amplitudes to zero and solve the positive frequency
problem by itself.

\medskip
\noindent {\bf Covariant Gauge Formalism}
\medskip
In the covariant gauge ladder (CGL) formalism, we decompose the $0^-$
Bethe--Salpeter
amplitudes into four independent spinor functions:
$$\chi_{_P}(P,q)=\gamma_{_5}\bigg\lbrack \chi_{_0}+\slash P \chi_{_1}+\slash q
\chi_{_2}+[\slash q,\slash P] \chi_{_3}\bigg\rbrack\; \; . \eqno (3.2)$$ In our
calculation, only $\chi_{_0}$ and $\chi_{_1}$ need to be retained, and the weak
coupling relation $\chi_{_1}=-\chi_{_0}/2m$ can be
employed to reduce the system of equations to a single equation for
$\chi_{_0}$.
In Euclidean variables, we have
$$\eqalign{\chi_{_0}(P,q)=&{4\alpha_s\over 3\pi^3} {(q^2+m^2)\over D}\; \int
d^4
k{\chi_{_0}\over (k-q)^2}\cr
&+{m^2g_2^2\over 4M_W^2}\; {1\over D}\; \biggl(q^2+{3M^2_B\over
4}+m^2\biggr)\int {d^4k\over (2\pi)^4}\; {\chi_{_0}\over (k-q)^2+M_H^2}\cr
&+{m^2g_2^2\over 4M_W^2}\; {F^I\over D}\;\biggl(q^2-{M^2_B\over
4}+m^2\biggr)
\int {d^4k\over (2\pi)^4}\; {\chi_{_0}\over (k-q)^2+M_Z^2}\cr}\; \; , \eqno
(3.3)$$
where $M_Z$=$M_W$=90GeV has been adopted since including weak boson mass
difference affects answers only in the third decimal place, $D\equiv
D(q^2,M_B^2,$cos$\theta)$=$(-q^2+{M_B^2\over 4}-m^2)^2$
$+M_B^2q^2\cos^2\theta,\; M_B^2=-P^2$ is the mass--squared of the
bound state, $\theta$ is the angle between $P$ and $q$ and $F^I=+1,-3$ for
iso--spin
$I=1,0$ respectively. As mentioned in the preceding section, the sign
difference in the last term between the $I=1$ and $I=0$ cases produces the
extra attraction and repulsion compared to the purely attractive gluon and
Higgs potentials. The method of solution is discussed in Ref. 3 in some detail.
The
resulting bound state mass values as a function of quark mass, $m,$ are shown
in
Fig. 2 in the curves labeled S1 and T1. In this calculation, the iso--singlet
(S1)
state's binding energy steadily
increases, but the bound state mass never sinks to zero in the fermion mass
range we have investigated. The more tightly bound
triplet state (T1) achieves zero mass when the quark mass reaches about 900
GeV. Above
750 GeV quark mass, the splitting between the different bound state masses
becomes quite pronounced. The $0^+$ and $1^-$ calculations proceed in a similar
fashion with the isosinglet being more tightly bound than the isotriplet.
  The details are not presented
here, since they do not illuminate the discussion.
\medskip
\noindent {\bf Salpeter Equation Formalism}
\medskip
Making the approximation 3.1b, integrating over $q_0$ in the Bethe--Salpeter
equation, and projecting out the $0^-$ channel amplitude, one obtains coupled
equations for the positive and negative frequency amplitudes, $\chi_{_\pm}
(q)$,
where $q\equiv\mid\vec q\mid$. The equations read
$$\eqalign{\biggl(E-2\omega \biggr) \chi_{_+} (q)&={1\over \pi q}\int
dq'q'\biggl(V_+ \chi_{_+}+V_- \chi_{_-}\biggr)\cr
\biggl(E+2\omega \biggr) \chi_{_-} (q)&={-1\over \pi q}\int
dq'q'\biggl(V_- \chi_{_+}+V_+ \chi_{_-}\biggr)\cr}
\eqno
(3.4)$$ where
$\omega=\sqrt{q^2+m^2}, E$ is the bound state eigenvalue and
$$\eqalign{V_\pm&=C_VQ_0(Z_V){2\omega\omega'\mp m^2\over\omega\omega'}\cr
&+C_{PS}\bigg\lbrace Q_0
(Z_{PS}){\omega\omega' \mp m^2\over\omega\omega'}
+\bigg\lbrack Z_{PS}Q_0(Z_{PS})-1\bigg\rbrack {qq'\over
\omega\omega'}\bigg\rbrace\cr
&+C_S\bigg\lbrace Q_0 (Z_S) {\omega\omega' \pm m^2\over\omega\omega'}
+\bigg\lbrack Z_{S}Q_0(Z_{S})-1\bigg\rbrack {qq'\over
\omega\omega'}\bigg\rbrace\cr}
\; \; .$$ The coefficients $C_V$
and variables $Z_V$ etc. are summarized in Table 1. Details on the method of
solution are given in Ref. 3.
\medskip
The $0^-$, ground state masses as a function of quark mass is shown for the
positive frequency only case in Fig. 2, curves S2 and T2. Again the splitting
between
isosinglet (S2)
and isotriplet (T2) bound state masses is striking for the heavy quark mass
region, and the iso--triplet system becomes ultrarelativistic in the region
above $m=750GeV$, producing a zero mass bound state at $m\simeq 1100 GeV$.
\medskip
The solution for the positive plus negative, fully coupled, system is much more
tightly bound as is seen in Fig. 2, curves S3 and T3, where a dramatic
departure between
iso--triplet (T3) and iso--singlet (S3) masses sets in already at $m=400GeV$,
and the
iso--triplet mass plunges to zero at $m=520GeV$. Figure 2 also shows the
feature of the instantaneous approximation that differs from the CGL
approximation, namely the turn--over (and eventual fall to zero) of the
iso--singlet bound
state mass as a function of the quark mass.
This effect does not yet appear in Fig. 2
for the iso-singlet positive-frequency-only case,
but it does occur at a quark mass above 1.20 TeV for this case as well.
\medskip
Focusing on the iso--triplet ground state mass values vs. quark mass curves in
Fig. 2
for the three solutions which were discussed above, we see that the general
features agree though the different
relativistic bound--state approximations produce different bound state mass
values for a given quark mass. One can
conclude that the Goldstone boson Higgs exchange plays an important role in the
calculations, that deep binding at or above 500 GeV quark mass occurs, and that
a dramatic weak iso--spin mass splitting is produced in the $Q\bar Q$ spectrum
of a degenerate, or nearly so,  $U,D$ doublet system.
\medskip
With the role of iso--singlet and iso--triplet reversed, the features
just outlined are present also in the $J^P=1^-$ and $0^+$ bound
states, and we display results of calculations for this system in Figs.
3 and 4. The principle distinction between $0^-$ on one hand and $1^-$ and
$0^+$ on the other is the iso--singlet/triplet splitting reversal and that
the latter two are typically less tightly bound.
\medskip
In the next section we take up the question of ultra--heavy quark decay
lifetime
and intra--doublet mass  splitting constraints.
\bigskip
\noindent {\bf 4. Constraints on Super--Heavy $U,D$ Quark Lifetimes and on
$U-D$ Splitting}
\medskip
Two related questions arise when one considers the phenomenology of
super--heavy
fermions. A crucial consideration for bound state physics is the comparison of
quark lifetime to the period of bound state motion. The situation for the heavy
top
quark
case has been considered by a number of authors,$^{12}$ and we apply the
argument of
Strassler and Peskin to the super--heavy, fourth generation in this
section. Related to the lifetime question is the $U-D$ mass--splitting
question. If the splitting is larger than the $W$--mass, then the higher mass
quark can decay to the lower mass quark by direct $W$--emission, presumably
with
no Cabbibo, Kobayashi--Maskawa (C--K--M) suppression. The width would
then be too broad to permit formation  of narrow bound states containing the
heavier quark. We review constraints imposed by measured $W$ and $Z$--boson
properties on the $U-D$ mass splitting, which we would like to take as
degenerate to a first approximation.
\medskip
\noindent {\bf Lifetime for decay to light quarks}
\medskip
The lighter of the $U,D$ doublet partners must decay to lighter generation
quarks, $q$,
and we assume that one transition is dominant. Calling the corresponding
C--K--M factor $V_{Qq}$, we have
$$\Gamma_{Q\to q}\sim  (180) MeV \bigg\lbrack {M_Q\over M_W}\bigg\rbrack^3\mid
V_{Qq}\mid^2\; \; ,\eqno (4.1)$$ where $M_Q>>M_q,M_W$ is assumed. To make a
conservative
estimate, we will include only the gluon binding in estimating the time needed
for bound state formation in a non--relativistic, weak binding approximation.
Including the strong binding, Yukawa interactions will only improve the
prospects for bound state formation. For pure QCD coupling,
the characteristic radius is
$a_0=({4\over 3}\alpha_s M_Q/2)^{-1}$ where ${4\over 3}\alpha_s$ is the
effective QCD coupling strength for the problem (with a characteristic velocity
$v_n\sim (4/3)\alpha_s/n$ for the $n^{th}$ radial excitation). Taking the ratio
of twice the diameter to the velocity as an $s$--state formation time,$^{12}$
one has
$t_{form}\sim {9n^3\over 2\alpha_s^2 M_Q}\; \; \; n=1,2\cdots$ and in the
ground state $t_{form}\sim 9/2\alpha_s^2M_Q$. The inverse Bohr radius is the
appropriate scale at which to evaluate $\alpha_s$ for the problem, and using
the
one-loop parameterization for $\alpha_s$
$$\alpha_s(\mu^2)=\; {\pi d\over \ell n(\mu^2/\Lambda^2_{QCD})}\; \;
,\eqno (4.2)$$ where $d=12/(33-2n_f)$,
along
with $a_0^{-1}=\biggl({2\over 3}\alpha_s (\mu^2)M_Q\biggr)$, leads to a
condition
on $\alpha_s(\mu^2)$:
$${2\over \pi d}\; \ell n\biggl({2\alpha_s(\mu^2)M_Q\over
3\Lambda_{QCD}}\biggr)=\alpha_s^{-1}(\mu^2)\; \;.\eqno (4.3)$$ We show a plot
of
$\alpha_s
(1/a_0)$ as a function of $M_Q/\Lambda_{QCD}$ in Fig. 5. As expected, since
the binding scale is much less than the value of $M_Q,\; \alpha\biggl({1\over
a_0}\biggr)$ is significantly larger than $\alpha(M_Q)$. The latter is also
displayed
for comparison purposes in Fig. 5.
\medskip
Equating the time of formation and the decay lifetime, one obtains an upper
bound on the value of $\mid V_{Qq}\mid$ for given values of $\Lambda_{QCD}$ and
$M_Q$.
Choosing $\Lambda_{QCD}=0.20 GeV$, and $\Lambda_{QCD}=0.10$ we show $\mid
V_{Qq}\mid_{max}$ vs. $M_Q$ in Fig. 6.
\medskip
Fig. 6 clearly shows the point, a surprise to us, that the constraints on
$V_{Qq}$ are rather weak, permitting for example $V_{Qq}\simeq 0.3\sim \sin
\theta_c\simeq
V_{us}$ for $M_Q\simeq 500 GeV$ with $\Lambda_{QCD}=0.20 GeV$. As mentioned
above, including the Higgs boson, Yukawa coupling effects will only improve
chances of
bound state formation and further weaken the constraints on $(V_{Qq})_{max}$.
This observation is encouraging for prospects of detecting narrow,
deeply bound states at higher energies.
\medskip
\noindent {\bf Mass splitting between $U$ and $D$}
\medskip
Turning to the question of mass splitting between the $U$ and $D$ quarks, we
use the definition$^{13}$
$$\sin^2\theta_W\equiv 1-{M_W^2\over M_Z^2}$$ and the parameterization of
radiative corrections in terms of the factor $\Delta r$ to write$^{14}$
$${M_W^2\over M_Z^2}\biggl(1-{M_W^2\over M_Z^2}\biggr)={\pi\alpha\over
\sqrt{2}G_F} {1\over M_Z^2} {1\over 1-\Delta r}\eqno (4.4)$$

The quantity $\Delta r$ can be expressed as
$$\Delta r=0.071-{\cos^2\theta_W\over \sin^2\theta_W}\delta\rho_H\; \; ,\eqno
(4.5)$$
where the heavy fermion factor $\delta\rho_H$ is given by
$$\delta\rho_H\simeq {3 G_F\over 8\pi^2\sqrt{2} }\biggl(M_t^2+\sum\limits_i
{C_i\over 3}\Delta M_i^2\biggr)\; \; ,\eqno (4.6)$$ and $$\Delta
M^2=M_1^2+M_2^2-{4M_1^2M_2^2\over M_1^2-M_2^2}\ell n {M_1\over M_2}\; \; .\eqno
(4.7)$$ The factor $C_i=3$ for super heavy quarks and $C_i=1$ for super--heavy
leptons.$^{15}$ The masses are designated as $M_t$ for the top quark and $M_1$
and
$M_2$ for super--heavy fermion generation members. Using the 1992 data
book$^{15}$
values for $M_W$ and $M_Z$ and the value $\sin^2\theta_W=0.230$, we find from
Eqs. 4.4--4.7 that the one standard deviation constraint is
$$\Delta M_F^2\equiv M_t^2+\sum\limits_i {C_i\over 3} \Delta M_i^2\leq (206
GeV)^2\; \; .\eqno (4.8)$$ For the case of one super heavy--quark generation
and a degenerate super--heavy lepton generation, the plot of the bound on $\mid
M_U-M_D\mid$
vs. $M_t$ is shown in Fig. 7. There is evidently room for considerable
mass--splitting in the bound (4.8), where $\mid M_U-M_D\mid <140 GeV$ for
$M_t=150 GeV$, for example. The dependence of the bound on the average value of
the heavy quark mass is very weak and can be ignored for our purposes.
\medskip
An independent bound can be obtained from $Z$ decay data.  The leptonic and
hadronic decay widths, for example, provide useful constraints. Defining the
quantity $\sin^2\bar\theta_W\equiv \kappa(M_Z)\sin^2\theta_W$, one has
approximately$^{16}$
$$\sin^2\bar\theta_W=\bigg\lbrack {1\over 2}-{1\over2}
\biggl(1-{4\pi\alpha(M_Z)\over \sqrt{2}G_FM_Z^2}\biggr)^{1/2}\bigg\rbrack
\bigg\lbrack 1-{3\alpha\over 16\pi\sin^2\theta_W(1-2\sin^2\theta_W)}
{\Delta M_F^2\over M_Z^2}\bigg\rbrack\; \; ,\eqno (4.9) $$ with $\kappa
(M_Z)=1+\biggl(3\alpha/16\pi\sin^4\theta_W\biggr)\biggl(\Delta
M_F^2/M_Z^2\biggr)$.  The value of $\alpha(M_Z)$ is obtained from electroweak
corrections to $\alpha$ that do not include the heavy quark effects. Using the
value $\Delta r=0.071$ to define $\alpha(M_Z)$ and adopting the first factor in
brackets in Eq. (4.9) as the value of $\sin^2\theta_W$ to be used in the
expression in the second bracket, we find
$$\sin^2\bar\theta_W=0.235\biggl(1-3.50\times 10^{-3}{\Delta M_F^2\over
M_Z^2}\biggr)\; \; .\eqno (4.10)$$ The expressions for the decay widths are
$$\Gamma(Z\to\ell^+\ell^-)={\sqrt{2}G_FM_Z^3\over
48\pi}\rho\bigg\lbrack1+(1-4\sin^2\bar\theta_W)^2\bigg\rbrack\eqno (4.11a)$$
and
$$\Gamma(Z\to {\rm hadrons})={\sqrt{2}G_FM_Z^3\over
8\pi}\rho\bigg\lbrack5-{28\over 3}\sin^2\bar\theta_W+{88\over
9}\sin^4\bar\theta_W-\delta_b\bigg\rbrack \times\biggl(1+{\alpha_s(M_Z)\over
\pi}\biggr)\;
\; .\eqno (4.11b)$$ The factors $\rho$ and $\delta_b$ in Eqs. 4.11 are given by
$$\rho=1+{3\alpha\over 16\pi\sin^2\theta_W\cos^2\theta_W}\; {\Delta M_F^2\over
M_Z^2}\; \; ,$$ and
$\delta_b=\bigg\lbrack\alpha(1+4\sin^2\theta_W/3-16\sin^4\theta_W/9)/8\pi
\sin^2\theta_W\cos^2\theta_W\bigg\rbrack
(M_t^2/M_Z^2)$, where $\sin^2\theta_W =0.235$ is used from Eq. (4.9)
onward to evaluate all of
the above expressions. The constraints on $\mid M_U-M_D\mid$ provided by the
1992 Particle Data Group values of $\Gamma(Z\to\ell^+\ell^-)$ and $\Gamma(Z\to$
hadrons) are illustrated in Fig. 7. We see again that significant
mass splitting in a hypothetical fourth generation of fermions is allowed. The
constraints on splittings are essentially independent of average mass.
\medskip
To conclude this section, we note that our assumption of degeneracy between
super--heavy generation fermions is consistent with the present bounds, but
that there could be shifts in our iso--triplet vs. iso--singlet $Q\bar Q$ bound
state masses due to ultra--heavy quark non--degeneracy effects. The pattern of
a mass--splitting between iso--singlet and iso--triplet states of a given spin
and parity would remain, however.
\bigskip
\noindent {\bf 5. Discussion and Conclusion}
\medskip
In the highly relativistic, strong binding regime, we have shown that the
contributions of the pseudoscalar, Goldstone--boson degrees of freedom to the
binding of ultra--heavy $Q\bar Q$ systems play a crucial role in determining
the bound state spectrum. Even though these pseudoscalar interactions become
negligible compared to the gluon and Higgs exchanges in the weak binding limit,
they are solely responsible for the large iso--singlet vs iso--triplet $Q\bar
Q$ bound state splitting which we found for ultra--heavy, fourth generation,
$Q=(U,D)$ quarks. Depending upon the specific approximation scheme used, we saw
(Figs. 2,3,4) that the splitting between $I=0$ and $I=1$ $Q\bar Q$ states
become
large for $M_Q$ in the range $0.4 TeV <M_Q<0.8 TeV.$
\medskip
Whether a splitting between bound states is large depends of course on the
decay width of the (quasi) bound states, and our estimates of $\mid V_{Qq}\mid$
in Sec. 4 and the analyses of decay widths of similar states into $q\bar q$,
$H,\ Z$ etc. in Ref. 17--21 indicate that the splittings between states will be
large compared to the decay widths for the $M_Q$ values in the range given
above. The states should, therefore, be clearly separated.
\medskip
For the production and decay of neutral $Q\bar Q$ (and lepton--antilepton)
ultra heavy bound states, Refs. 2 and 17--21 provide an encouraging picture
of the production and decay signals for such states. The binding indicated by
our relativistic, bound state calculations is much stronger than that
considered in Refs. 17--18 (gluon non--relativistic potential model only) or
3, 19--21 (Higgs non--relativistic model or Higgs plus gluon non--relativistic
potential model), so the prospects for detecting new heavy bound states, should
a fourth generation exist, are made even brighter by our results.
\medskip
Avenues for further work include the detailed phenomenonlogy of the
iso--multiplet system, in particular the possible production and decay of the
charged state, and the application of these techniques to TeV scale
supersymmetry bound state physics. We plan to address these issues in the
future.
\bigskip

\noindent
{\bf Acknowledgements}
\bigskip

P.J. and D.M. thank H. Munczek and J. Ralston
for useful discussions. This work was performed in part at Ames
Laboratory under Contract No. W-7405-Eng-82 with the U.S.
Department of Energy and was supported in part by the U.S. Department
of Energy Grants Nos. DE-FG02-85ER40214 and DE-FG02-87ER40371, Division
of High Energy and Nuclear Physics. One of the authors (A.J.S.)
acknowledges financial support from a grant to Iowa State University
from the U.S. Department of Education, Graduate Assistance in Areas
of National Need Program.

\bigskip

\centerline{\bf   APPENDIX A}
\bigskip
We describe here the substraction procedure used in regulating the B--S
equation.
Referring to Eqs. (2.1) and (2.2) in the text, we see that the direct,
$P^2$--pole terms
in each equation are independent of $q$. All of the
other terms in the integrands depend on $k-q$ and they decrease like
$(k-q)^{-2}\sim q^{-2}$ as $\mid q\mid\to\infty$ for fixed $k$. Therefore, if
the integrals exists, one finds that for fixed $P$, taking (2.1) in particular,
$$\eqalign{{\lim\atop\mid q\mid\to\infty} S^{-1}(q_+)\chi_{_{UU}}(q,P)S^{-1}
(q_-)&\to
{g_2^2\over 4} {m^2\over M_W^2}
{i\over P^2-M_Z^2}\gamma_5 Tr \biggl(\int {d^4k\over
(2\pi)^4}\chi_{_{UU}}(k,P)\gamma_5\biggr)\cr
&+{g^2_2\over 4}{m^2\over M^2_W} {i\over
P^2-M_H^2}Tr\biggl(\int {d^4k\over
(2\pi)^4}\chi_{_{UU}}(k,P)\biggr)\cr}
\; .\eqno (A.1)$$ If $S^{-1}(q_+)_{\mid
q\mid\to\infty}\to\slash q$ as is the case for the perturbative propagators
which we use in
this paper, then the fixed $P$, large $q$ limit gives
$$\chi(q,P)\to {1\over q^2}\; \; .$$ In this case, the integral
$\int d^4k\chi(k,P)$ does not exist and the integral equation has no solutions,
in accordance with direct numerical studies which we have carried out.
\medskip
If the $q$--independent direct $(p)$--channel pole terms are absent, however,
then the $q^{-2}$ dependence of the integrands of all of the other terms in the
large $q^2$ limit means that $\chi(q,P)_{\mid q\mid\to\infty}\to
q^{-4}$ behavior is consistent with the existence of solutions, and we find
that this is indeed the case.
\medskip
To render finite the equations (2.1), (2.2) and their counterparts with $U$ and
$D$ interchanged, it is sufficient to subtract each equation at a fixed $q$,
say $q=0$, because this subtraction removes the $q$--independent, divergent
term. To make this subtraction systematic, one regulates ({\it i.e.}
cuts--off) the integrals, makes the subtraction, and then removes the
regularization.
The subtracted equation, is as follows:
$$\eqalign{S^{-1}(q_+)\chi_{_{UU}}(q,P)&S^{-1}(q_-)
-S^{-1}\biggl({P\over 2}\biggr)\chi_{_{UU}}(0,P)S^{-1}\biggl({-P\over 2}\biggr)
\cr
&=\int {d^4k\over
(2\pi)^4}\bigg\lbrack -\gamma_\mu\chi_{_{UU}}(k,P)\gamma_\nu
G^{\mu\nu}(k-q)+{g_2^2\over 4}{m^2\over M_W^2}\biggl(\chi_{_{UU}}(k,P){i\over
(k-q)^2-M_H^2}\cr
&+\gamma_5\chi_{_{UU}}(k,P)\gamma_5{i\over
(k-q)^2-M_Z^2}+2\gamma_5\chi_{_{DD}}(k,P)\gamma_5{i\over
(k-q)^2-M_W^2}\biggr)\bigg\rbrack\cr
&-\int {d^4k\over
(2\pi)^4}\bigg\lbrack -\gamma_\mu\chi_{_{UU}}(k,P)\gamma_\nu
G^{\mu\nu}(k)-{g_2^2m^2\over 4M_W^2}\biggl(\chi_{_{UU}}(k,P){i\over
k^2-M_H^2}\cr
&+\gamma_5\chi_{_{UU}}(k,P)\gamma_5{i\over
k^2-M_Z^2}+2\gamma_5\chi_{_{DD}}(k,P)\gamma_5{i\over
k^2-M_W^2}\biggr)\bigg\rbrack\; \; .\cr}\eqno (A.2)$$ If the first
term on the left hand side of (A.2) is equal to the first integral on the right
hand side for each $q$ ({\it i.e.}, a solution for each $q$ is found), then the
companion terms, where $q=0$, on each side are guaranteed to be equal and the
result is an eigenvalue equation for $P$ and $q$.
This equation is, of course,
just the original equation without the offending, divergent term and we use
this form as our regulated, finite Bethe--Salpeter equation which is to be
projected into states of definite quantum numbers.

\bigskip
\centerline{\bf   APPENDIX B}
\bigskip
\noindent {\bf Running of the Yukawa Coupling -- The Landau Pole}
\medskip
In this appendix we briefly discuss the effects that arise because of the
momentum dependence, or running, of the Yukawa couplings. We work at the
one--loop level to see where the Landau pole occurs when the quark masses are
in
the deep binding region that we probe, $0.5 TeV\applt M_Q\applt 1.0 TeV$.
\medskip
The one--loop renormalization group equations for the gluon and Yukawa
couplings, $g_s$ and $g_Y$, without the weak gauge interaction effects,
read$^{22}$
$${d\over dt}g_s=-{1\over 4\pi^2 d}g_s^3\; \; ,\eqno (B.1a)$$ and
$${d\over dt}g_Y={1\over 4\pi^2} \biggl({9\over 8} g_Y^3-2g_s^2
g_Y\biggr)\; \; ,\eqno (B.1b)$$ where $d=12/(33-4n_g)$, $n_g$ is the
number of generations, and $t={1\over 2}\ell n\biggl({q^2\over \mu^2}\biggr)$
with $\mu$ an arbitrary renormalization scale.
\medskip
Parameterizing the solution to (B.1a) in the standard fashion,$^{23}$
$$g_s^2(q^2)={4\pi^2d\over \ell n \biggl(q^2/\Lambda_{QCD}^2\biggr)}\equiv
{2\pi^2 d\over t}$$ where $\Lambda_{QCD}\simeq 0.2$ GeV and $\mu=\Lambda_{QCD}$
is adopted, we can rewrite (A.1b) as
$${d g_{_Y}\over dt}={9\over 32\pi^2}g_{_Y}^3 -{d\over t}\; g_{_Y}\; \; .\eqno
(B.2)$$ The solution to (B.2) can be written in terms of an integration
constant
$C$, to be fixed by the boundary condition:
$$g_Y^2(t)={1\over Ct^{2d}-{9\over 16\pi^2}\; \; {t\over 1-2d}}\; .\eqno
(B.3)$$

Checking several limiting cases of (B.3), we see that
$$g_Y^2(t)={g^2_Y (0)\over 1-{9\over 16\pi^2} g^2_Y(0) t}$$ when
$\alpha_{_{QCD}}=0\; d=0 $, as it should. The characteristic  Landau pole
appears at $t={16\pi^2\over 9}\; {1\over g_{_{Y}}^2(0)}$.  In the
weak Yukawa coupling limit, which is the light quark case, where only the
$g_s^2$ effect is kept,
$$g_Y^2(t)={C^{-1}\over t^{2d}}\; \; ,$$ so the quark mass, proportional
to $g_Y$ and the weak scale $v$, behaves as
$$M_Q\sim g_Y (t){v\over \sqrt{2}}\to \; {1\over t^d},\;
t\to\infty\; \; ,$$ which is the well known leading--log, asymptotic
behavior of the light quark masses in QCD.$^{24}$
\medskip
The constant $C$ in Eq. (B.3) can be fixed in terms of the input quark mass,
$M_Q$, by requiring the condition$^{25}$
$$M_Q^2=g_Y^2 (2M_Q){v^2\over 2}\; \; ,\eqno (B.4)$$ for example. The
requirement (B.4) can be solved for $C$ to give
$$C=t_Q^{-2d}\biggl({v^2\over 2M_Q^2}-{9\over 16\pi^2}\; {1\over 2d-1}
t_Q\biggr)\; \; , \eqno (B.5)$$ with $$t_Q\equiv \ell n {2M_Q\over
\Lambda_{QCD}}\; \; .$$
The expression B.3 with $C$ evaluated by (B.5) can now be used to assess the
large $\mid q\mid>M_Q$, behavior of $g_{_Y}(t)$.
We can use the results (B.5) and (B.3) to determine $C_Q$ and $q_{pole}$, the
location of the singularity for various $M_Q$ values. There is no longer a
pole at large $t$, but the coupling still blows up at large $t$.
Interesting values are:
$M_Q=0.5 TeV$, where $C_Q=-0.282$ and $q_{pole}=13 TeV$; and
$M_Q=1.0 TeV$, where $C_Q=-0.288$ and $q_{pole}=3.5 TeV$.
The $t=0$ pole in (B.3) is an artifact of the
standard parameterization of $q_s$, which is only applicable when $\mid q\mid
>>\Lambda_{QCD}$. $g_{_Y}$ should be sensibly constant for small momentum
transfers.
\medskip
The momentum values that are important in the wave function in the
Bethe--Salpeter integretion are of the order of the inverse Bohr radius, as a
rough rule
of
thumb. Only in those cases when $M_Q\appgt 1TeV$ {\it and} in the limit as the
bound state mass plunges to zero does the $g_Y$ singularity come near to the
relevant range of integration in the Bethe--Salpeter equation. Consequently,
the binding energies as a function of $M_Q$ should not be substantially
affected by the $g_Y$ singularity below the point where the bound state masses
fall to zero, even for the $M_Q\appgt 1TeV$ cases.

\vfill\eject
\centerline{\bf Figure Captions}
\noindent Fig. 1: Bethe-Salpeter equations for the $U\bar D$
and $U\bar U$ bound states.

\noindent Fig. 2: Isosinglet and isotriplet masses for mesons in the
$0^-$ channel. Curves S1 (isosinglet) and T1 (isotriplet) are the results
from covariant gauge formalism. S2 (isosinglet) and T2 (isotriplet) are the
results from Salpeter's equation using only positive frequency
components. S3 (isosinglet) and T3 (isotriplet) are the results from
Salpeter's equation using both positive and negative frequency components.

\noindent Fig. 3: Isosinglet and isotriplet masses for mesons in the
$1^-$ channel. Solid curve (isosinglet) and dotted curve (isotriplet)
are the results from covariant gauge formalism.
Short dash curve (isosinglet) and long dash curve (isotriplet) are the
results from Salpeter's equation using only positive frequency
components. Short dash-dot curve (isosinglet) and long dash-dot (isotriplet)
are the results from
Salpeter's equation using both positive and negative frequency components.

\noindent Fig. 4: Isosinglet and isotriplet masses for mesons in the
$0^+$ channel. Solid curve (isosinglet) and dotted curve (isotriplet)
are the results from covariant gauge formalism.
Short dash curve (isosinglet) and long dash curve (isotriplet) are the
results from Salpeter's equation using only positive frequency
components. Dash-dot curve is the isosinglet result from Salpeter's
equation with positive and negative frequency components. The
isotriplet result for this case is almost identical to the
corresponding result obtained by including only the positive frequency
components in the Salpeter's equation.

\noindent Fig. 5: The value of $\alpha_s$ (solid curve) relevant for the bound
state Bethe-Salpeter equation, ignoring higgs interaction,
 as a function of the quark mass.
The dashed curve shows $\alpha_s(M_Q^2)$ which is significantly
smaller than the value of $\alpha_s$ that should be used for bound state
calculations.

\noindent Fig. 6: The maximum value of the CKM matrix element $|V_{Qq}|$
allowed in order for the fourth generation quarks $Q$ to form a meson
bound state, $q$ being a lower generation quark. This constraint
is derived by including only QCD effects. Inclusion of the Yukawa
coupling will allow even higher values of $|V_{Qq}|$.

\noindent Fig. 7: The maximum allowed value for the fourth generation
quark mass difference $|M_U-M_D|$ as a function of the top quark
mass $M_t$. The three curves represent the constraints using
$\rho$ parameter (solid curve), $\Gamma(Z\rightarrow hadrons)$
(short dashed curve) and $\Gamma(Z\rightarrow l^+l^-)$ (long dashed
curve).

\vfill\eject
\centerline{\bf References and Footnotes}
\bigskip
\item{1.} Several recent examples are: C. Hill, M. Luty and E. Paschos, {\it
Phys. Rev.} {\bf D43}, 3011 (1991); S. King, {\it Phys. Lett. B.} {\bf 281},
295 (1992). A recent discussion of ultra heavy quarks and heavy Higgs dynamical
effects in a four generation setting is given in T. Truong, {\it Phys. Rev.
Lett.} {\bf 70}, 888 (1993).
\bigskip
\item{2.} H. Inazawa and T. Morii, {\it Phys. Lett.} {\bf 203B}, 279 (1988).
This paper introduces a Yukawa potential, Schrodinger equation version of
strong Higgs--boson coupling to fermions.
\bigskip
\item{3.} P. Jain, A. Sommerer, D. McKay, J. Spence, J. Vary and B.--L. Young,
{\it Phys.
Rev. D} {\bf 46}, 4029 (1992). The relativistic bound state formalism used in
the present paper is explained in this reference.
\bigskip
\item{4.} M. Chanowitz, M. Furman and I. Hinchliffe, {\it Phys. Lett.} {\bf
78B}, 285 (1978); M. Chanowitz, M. Furman and I. Hinchliffe, {\it Nucl. Phys.}
{\bf B153}, 402 (1979).
\bigskip
\item{5.} M. Veltman, {\it Nucl. Phys.} {\bf B123}, 89 (1977); P.Q. Hung, {\it
Phys. Rev. Lett.} {\bf 42}, 873 (1979); H. Politzer and S. Wolfram, {\it Phys.
Lett.} {\bf 82B}, 242 (1979); {\bf 83B}, 421 (E) (1979).
\bigskip
\item{6.} We use the term ultra--heavy to distinguish fourth and
higher generation fermions from the $c,b,$ and $t$ quarks, generally referred
to as heavy quarks.
\bigskip
\item{7.} We refer
to the $SU(3)\times
SU_L(2)\times U(1)$ model with three generations of fermions and one scalar
doublet as the standard model.
\bigskip
\item{8.} In the unitary gauge, where the Goldstone bosons do not appear, their
role is played by the longitudinal degrees of freedom of the $W^\pm$ and $Z$ at
the perturbative level.
\bigskip
\item{9.} These considerations were not included in Ref. 3.
\bigskip
\item{10.}  Zero mass bound in strong binding were discussed by J. Goldstone,
{\it Phys. Rev.} {\bf 91}, 1516 (1953). For a recent review that includes
discussion of this effect, see N. Seto, {\it Suppl. Prog. Theor. Phys.} {\bf
95}, 25 (1988).
\bigskip
\item{11.} Vacuum phase transition and tachyonic bound states are studied in
scalar theories by R. Haymaker, {\it Phys. Rev.} {\bf D13}, 968 (1976); {\it
ibid} {\bf D16}, 1211 (1977).
\bigskip
\item{12.} I.I. Bigi {\it et al.}, {\it Phys. Lett. B} {\bf 181}, 157 (1986);
J.H. K\"uhn and P.M. Zerwas, {\it Phys. Rep.} {\bf 167}, 321 (1988);
V.S. Fadin and V.A. Khoze, {\it Pis'mas Zh. Eksp. Theor. Fiz.} {\bf
46}, 417 (1987) [{\it JETP Lett.} {\bf 46}, 525 (1987)];
M. Strassler and M. Peskin, {\it Phys. Rev. D.} {\bf 43}, 1500
(1991).
\bigskip
\item{13.} A. Sirlin, {\it Phys. Rev. D} {\bf 22}, 971 (1980); W. Marciano and
A. Sirlin, {\it Phys. Rev. D} {\bf 22}, 2695 (1980).
\bigskip
\item{14.} See, for example the discussion of G. Altarelli, Carg\`ese Summer
Institute on Particle Physics 18 July--4 August, 1989 (CERN--TH.5590/89).
\bigskip
\item{15.} Review of Particle Properties, {\it Physical Review D} {\bf 45},
Number 11, Part II (1992).
\bigskip
\item{16.} See, for example, R. Peccei, {\it Mod. Phys. Lett. A} {\bf 5}, 1001
(1990).
\bigskip
\item{17.} V. Barger {\it et al.} {\it Phys. Rev. Lett.} {\bf 57}, 1672 (1986).
\bigskip
\item{18.} V. Barger {\it et al.} {\it Phys. Rev.} {\bf D35}, 3366 (1987).
\bigskip
\item{19.} H. Inazawa and T. Morii, {\it Z. Phys. C.} {\bf 42}, 563 (1989).
\bigskip
\item{20.} H. Inazawa, T. Morii and S. Tanaka, {\it Z. Phys. C.} {\bf 43}, 569
(1989).
\bigskip
\item{21.} H. Inazawa, T. Morii and J. Morishita, {\it Z. Phys. C} {\bf 46},
273 (1990).
\bigskip
\item{22.} C. Hill, {\it Phys. Rev. D} {\bf 24}, 691 (1981). The general
framework for Higgs Yukawa couplings was laid out in in T.P. Cheng, E. Eichten
and L.F. Li, {\it Phys. Rev.D} {\bf 9}, 2259 (1974).
\bigskip
\item{23.} This choice has the advantage of the familiarity. The results that
follow are meant to apply at $t>M_Q>>\Lambda_{QCD}$, however. For an
alternative formulation of the solution see, Hill, Ref. 17.
\bigskip
\item{24.} H.D. Politzer, {\it Nucl. Phys.} {\bf B117}, 397 (1976).
\bigskip
\item{25.} H. Georgi and H.D. Politzer, {\it Phys. Rev. D} {\bf 14}, 1829
(1976).
\vfill\eject
\newskip\hidevskip

\def\tablerule{\noalign{\hrule height 1pt}}

\centerline{\bf Table 1}

$$\vbox{\tabskip=0pt \offinterlineskip
        \halign{\strut#&\vrule width 1pt \tabskip=2em plus .4em#&
        \lft{#}&\vrule width 1pt#&\ctr{#}&\vrule width 1pt#&
        \ctr{#}&\vrule width 1pt#&\ctr{#}&\vrule width 1pt#\tabskip=0pt\cr

&\omit & &\omit &$V$ &\omit &$PS$&\omit &$S$ &\omit\cr\tablerule
& & & & & & & & &\cr
& &$C, I=0$ & &$+{4\over 3}\alpha_s$& &${-3\over 4\pi}$ ${m^2\over v_0^2}$ &
&${1\over 4\pi}$ ${m^2\over v_0^2}$ &\cr
& & & & & & & & &\cr\tablerule
& & & & & & & & &\cr
& &$C, I=1$& & $+{4\over 3}\alpha_s$& &$+{1\over 4\pi}$ ${m^2\over v_0^2}$ &
&$+{1\over 4\pi}$ ${m^2\over v_0^2}$ &\cr
& & & & & & & & &\cr\tablerule
& & & & & & & & &\cr
& &$Z$ & &${q^2+q^{'2}\over 2 qq'}$& &${q^2+q^{'2}+M_Z^2\over 2qq'}$&
&${q^2+q^{'2}+M_H^2\over 2 qq'}$ &\cr
& & & & & & & & &\cr\tablerule}}$$
\vskip .5in
\noindent {\bf Table 1}
\bigskip
\noindent Coefficients for the $0^-$ channel Salpeter equation kernels, Eq.
(3.4) and below. Positive signs indicate attractive interactions and the
negative
sign a repulsive interation.

\bye